\UseRawInputEncoding 
\documentclass[superscriptaddress,aps,prl,reprint,twocolumn,amsmath,amssymb,showpacs,]{revtex4-2}
\usepackage{graphicx}  
\usepackage{mathptmx}
\usepackage{epstopdf}
\usepackage{dcolumn}   
\usepackage{bm}        
\usepackage{amssymb}   
\usepackage{amsmath}   
\usepackage{amsthm}
\usepackage{chemarrow}
\usepackage{color}
\usepackage{mathrsfs}
\usepackage{float}
\usepackage{bbold}
\usepackage{bbm}
\usepackage{braket}
\usepackage[a4paper,colorlinks=true,
linkcolor=blue,citecolor=blue,
pdfauthor={ },
pdftitle={ },
pdfsubject={ },
pdfkeywords={ }]{hyperref}

\theoremstyle{plain}
\newtheorem*{theorem*}{Theorem}

\usepackage{verbatim}

\begin{document}

\title{Collision-induced spin noise}

\author{Shiming Song}

\affiliation{
Hefei National Laboratory for Physical Sciences at the Microscale and Department of Modern Physics, University of Science and Technology of China, Hefei 230026, China}
\affiliation{
CAS Key Laboratory of Microscale Magnetic Resonance, University of Science and Technology of China, Hefei, Anhui 230026, China}
\affiliation{
Synergetic Innovation Center of Quantum Information and Quantum Physics, University of Science and Technology of China, Hefei, Anhui 230026, China}

\author{Min Jiang}
\email{dxjm@ustc.edu.cn}
\affiliation{
Hefei National Laboratory for Physical Sciences at the Microscale and Department of Modern Physics, University of Science and Technology of China, Hefei 230026, China}
\affiliation{
CAS Key Laboratory of Microscale Magnetic Resonance, University of Science and Technology of China, Hefei, Anhui 230026, China}
\affiliation{
Synergetic Innovation Center of Quantum Information and Quantum Physics, University of Science and Technology of China, Hefei, Anhui 230026, China}

\author{Yushu Qin}

\affiliation{
Hefei National Laboratory for Physical Sciences at the Microscale and Department of Modern Physics, University of Science and Technology of China, Hefei 230026, China}
\affiliation{
CAS Key Laboratory of Microscale Magnetic Resonance, University of Science and Technology of China, Hefei, Anhui 230026, China}
\affiliation{
Synergetic Innovation Center of Quantum Information and Quantum Physics, University of Science and Technology of China, Hefei, Anhui 230026, China}

\author{Yu Tong}

\affiliation{
Hefei National Laboratory for Physical Sciences at the Microscale and Department of Modern Physics, University of Science and Technology of China, Hefei 230026, China}
\affiliation{
CAS Key Laboratory of Microscale Magnetic Resonance, University of Science and Technology of China, Hefei, Anhui 230026, China}
\affiliation{
Synergetic Innovation Center of Quantum Information and Quantum Physics, University of Science and Technology of China, Hefei, Anhui 230026, China}

\author{Wenzhe Zhang}

\affiliation{
Hefei National Laboratory for Physical Sciences at the Microscale and Department of Modern Physics, University of Science and Technology of China, Hefei 230026, China}
\affiliation{
CAS Key Laboratory of Microscale Magnetic Resonance, University of Science and Technology of China, Hefei, Anhui 230026, China}
\affiliation{
Synergetic Innovation Center of Quantum Information and Quantum Physics, University of Science and Technology of China, Hefei, Anhui 230026, China}

\author{Xi Qin}
\affiliation{
Hefei National Laboratory for Physical Sciences at the Microscale and Department of Modern Physics, University of Science and Technology of China, Hefei 230026, China}
\affiliation{
CAS Key Laboratory of Microscale Magnetic Resonance, University of Science and Technology of China, Hefei, Anhui 230026, China}
\affiliation{
Synergetic Innovation Center of Quantum Information and Quantum Physics, University of Science and Technology of China, Hefei, Anhui 230026, China}

\author{\mbox{Ren-Bao Liu}}
\affiliation{
Department of Physics, Centre for Quantum Coherence, and The Hong Kong Institute of Quantum Information Science and Technology, The Chinese University of Hong Kong, Shatin, New Territories, Hong Kong, China}

\author{\mbox{Xinhua Peng}}
\email{xhpeng@ustc.edu.cn}
\affiliation{
Hefei National Laboratory for Physical Sciences at the Microscale and Department of Modern Physics, University of Science and Technology of China, Hefei 230026, China}
\affiliation{
CAS Key Laboratory of Microscale Magnetic Resonance, University of Science and Technology of China, Hefei, Anhui 230026, China}
\affiliation{
Synergetic Innovation Center of Quantum Information and Quantum Physics, University of Science and Technology of China, Hefei, Anhui 230026, China}

\begin{abstract}
Collision phenomena are ubiquitous and of importance in determining the microscopic structures and intermolecular interactions of atoms and molecules.
The existing approaches are mostly based on atomic or molecular scatterings, which are hindered by the inconvenience of using ultra-high vacuum and low temperature systems.
Here we demonstrate a new spin-noise spectroscopic approach by measuring optical polarization rotation noise of the probe light, which operates with simple apparatus and ambient conditions. 
Our approach features tens of gigahertz bandwidth and one part-per-million resolution,
outperforming existing spin-noise techniques.
Enabled by the new technique, we observe the collision-induced spin noise of alkali atoms, and precisely determine key collision parameters, such as collision diameter, well depth, and dominant interaction type.
Our work provides a new tool to study a broad range of collision phenomena under ambient conditions.
\end{abstract}

\maketitle

\emph{Introduction}.-Collisions are ubiquitous in physics, chemistry and thermodynamics.
The analysis of collision phenomena plays important roles in determining the microscopic structures \cite{goldberger2004collision,xiong2019small}, interactions \cite{frommhold2006collision,bartocci2014intermolecular,karman20182} and energy stability \cite{robinson1960frequency,herman1968rare,bernheim1969effects,vanier1974relaxation,sortais2000cold,gong2008nonlinear} of atoms and molecules.
Collision phenomena have been studied by a large variety of techniques,
such as scattering experiments~\cite{goldberger2004collision,bartocci2014intermolecular,xiong2019small}, absorption spectra~\cite{frommhold2006collision,karman20182,fauchez2020sensitive} and nuclear magnetic resonance~\cite{jameson1991gas,chen2004new}.
Among these techniques, scattering experiments using atomic or molecular beams are mostly used.
However, such scattering methods usually require high vacuum degree less than 10$^{-6}$~bar and low temperature as low as tens of kelvin \cite{bartocci2014intermolecular},
which place severe limits in their realistic applications.
It remains challenging to, for example, study atomic collisions under relatively high pressures and temperatures, such as 1~bar and 400~K respectively,
which are the usual conditions for many important applications such as atomic vapor magnetometers \cite{budker2007optical,jiang2018experimental,jiang2019magnetic,jiang2020interference,jiang2021floquet} and electrometer \cite{jing2020atomic}.
Therefore, it is highly desirable to develop new techniques that can investigate collision phenomena in wide-ranging experimental conditions.

Colliding particles may exhibit new properties, such as energy level shifts \cite{robinson1960frequency,herman1968rare,bernheim1969effects,vanier1974relaxation}, collision-induced absorption \cite{frommhold2006collision,karman20182,fauchez2020sensitive} and interatomic entanglement \cite{blum2016entanglement,luo2017deterministic,kong2020measurement,mouloudakis2021spin}.
Recently, collisions between alkali atoms and chemically inert atoms or molecules have attracted considerable attention in the areas of frequency standards \cite{robinson1960frequency,vanier1974relaxation}, metrology \cite{budker2007optical,jiang2021floquet,jiang2019magnetic,jiang2020interference,jiang2018experimental,jing2020atomic}, and quantum information~\cite{kong2020measurement,mouloudakis2021spin}.
Recently,
direct measurement of quantum spin noise in thermodynamic equilibrium by optical rotation is becoming a mainstream approach for non-perturbative studies of energy structures \cite{aleksandrov1981magnetic,crooker2004spectroscopy,oestreich2005spin,mihaila2006quantitative,muller2008spin,li2013nonequilibrium,zapasskii2013spin,glasenapp2014spin,cronenberger2015atomic,yang2014two,sinitsyn2016theory,tang2020spin}, spatial properties \cite{lucivero2017correlation,kozlov2018spin,cronenberger2019spatiotemporal} and correlated states \cite{eckert2008quantum,bruun2009probing,chen2014faraday,li2016universality} in diverse systems, such as alkali atomic vapors \cite{aleksandrov1981magnetic,crooker2004spectroscopy,zapasskii2013optical,roy2015cross}. 
Collision phenomena can be investigated by measuring the energy spectrum of colliding particles via spin-noise techniques.
However, when it is used to measure collision-induced atomic energy shifts,
this places high demands both for measurement bandwidth and for spectral resolution.
Specifically,
direct observation of collisional energy shifts of Zeeman sublevels is much more difficult than that of inter-hyperfine levels.
For example, the shifts of Zeeman sublevels at a magnetic field of 1 G are seven order of magnitude smaller than that of the inter-hyperfine levels~\cite{herman1968rare}. 
Accordingly, to detect collisional shifts of microwave inter-hyperfine levels, the gigahertz bandwidth is required.
The other requirement is high spectral resolution, because collisional shifts are still only 1-10~part-per-million for inter-hyperfine levels~\cite{robinson1960frequency,bernheim1969effects,vanier1974relaxation}.
Unfortunately, existing spin-noise techniques based on optical-rotation approach suffer from a trade-off between bandwidth and spectral resolution~\cite{berski2013ultrahigh,zapasskii2013spin,cronenberger2016quantum,sinitsyn2016theory}.
The best spectral resolution is only several hundred kilohertz at the gigahertz frequency range~\cite{cronenberger2016quantum}.

In this Letter, we demonstrate a first realization of measuring collision phenomena by developing and applying spin-noise spectroscopy that has gigahertz bandwidth and part-per-million spectral resolution.
Collision-induced spin noises arising from inter-hyperfine levels are detected by the optical rotation.
The noise spectra are analyzed to reveal the collisional energy shifts.
We apply our technique to measure spin noise of alkali atoms colliding with a variety of chemically inert molecules,
and obtain key collision parameters,
including the collision diameter, the depth of the potential well, and the dominant interaction type.
The results are in good agreement with theory.
Besides, the relative strength of spin noise scales inversely with the probe volume \cite{crooker2004spectroscopy,zapasskii2013optical,sinitsyn2016theory},
thus our method appears favourable for measuring the collision dynamics at mesoscopic and even microscopic scales \cite{laliotis2014casimir,budker2019extreme}.
Our technique can be applied to a broad range of spin noise-based applications, such as determining structural information \cite{crooker2004spectroscopy,cronenberger2015atomic} and fundamental precision limits of microwave devices \cite{pedrozo2020entanglement}.

\emph{Spin noise as a probe of collision}.-We consider the case of binary collision, where two particles interact through an effective potential depending on their relative distance.
For an alkali atom in the binary collision with a colliding molecule, the wave function of the valence electron is perturbed and thus the energy level undergoes a shift.
According to the statistical theorem, the mean energy shift of the alkali atomic ensemble equals the statistical mean of all perturbations under every general condition \cite{robinson1960frequency}. 
Therefore, the collisional frequency shift can be expressed as (here we assume $h=1$) \cite{robinson1960frequency}
\begin{equation}
\begin{aligned}\label{shift1}
\nu_{\rm{shift}}=4\pi n\int\delta E(r)  e^{-U(r)/k_BT}r^2dr,\\
\end{aligned}
\end{equation}
where $r$ is the distance between the collision pair,
$U(r)$ is the effective potential between the collision pair,
$\delta E(r)$ is the energy perturbation on the alkali atom from collisions,
$n$ is the density of the colliding molecules, and $T$ is the temperature.

\begin{figure}[htb]  
	\makeatletter
	\def\@captype{figure}
	\makeatother
	\includegraphics[scale=0.4]{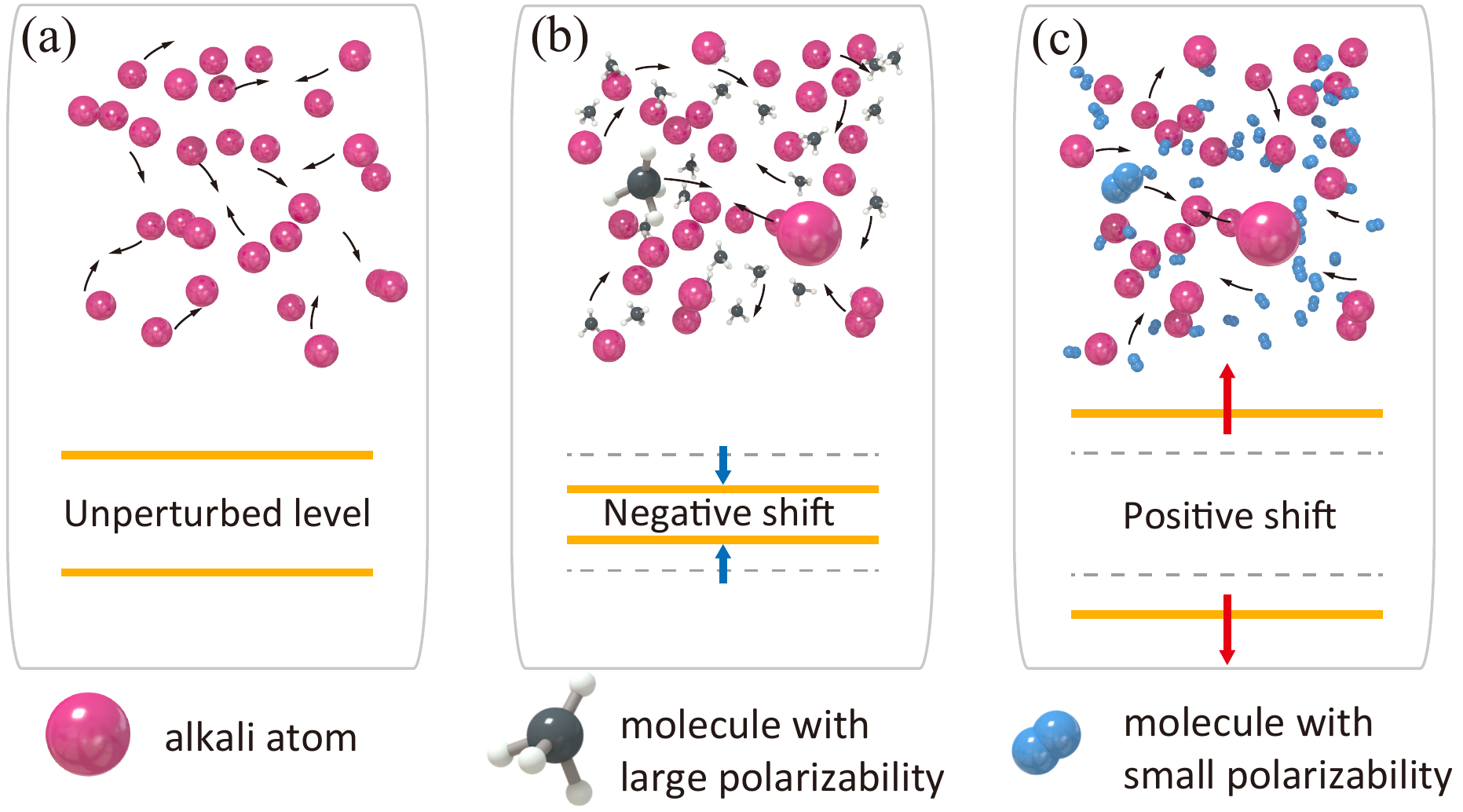}
	\caption{(color online).
	Collision-induced spin energy shifts.
	(a) Unperturbed energy level of alkali atoms, such as rubidium atoms.
    (b) Negative energy shift of alkali atoms colliding with molecules with large electric polarizability, such as Rb$-$CH$_4$ collision.
    (c) Positive energy shift of alkali atoms colliding with molecules with small electric polarizability, such as Rb$-$N$_2$ collision. }
	\label{figure-0}
\end{figure}

The intermolecular potential and the energy perturbation can be simplified as the summation of a long-range van der Waals attraction and a short-range Pauli repulsion, such as the Lennard-Jones potential~\cite{kaplan2006intermolecular}
\begin{equation}
\begin{aligned}
    \label{shift2}
    U(r)=4\epsilon_1[(\sigma_1/r)^{12}-(\sigma_1/r)^{6}],\\
    \delta E(r)=4\epsilon_2[(\sigma_2/r)^{12}-(\sigma_2/r)^{6}].
    \end{aligned}
\end{equation}
Here $\sigma_1$ is the node where the potential $U(r)$ is zero, and is called the collision diameter.
$\epsilon_1$ is the well depth.
Similarly, $\delta E(r)$ is zero at $r=\sigma_2$ and its minimum is $-\epsilon_2$. 
As an example of collisional shifts shown in Fig.~\ref{figure-0}, the hyperfine splitting of the alkali atom arises from the Fermi-contact interaction of the valence electron and nucleus, and is proportional to the valence electron density at the nucleus \cite{bernheim1969effects}. 
Figure~\ref{figure-0}(a) shows such an alkali ensemble and the corresponding unperturbed energy levels (we adopt two of these levels as an example).
Here we neglect the shifts induced by collisions between alkali atoms themselves, which are usually less than 10$^{-6}$~Hz \cite{sortais2000cold}.
The attractive force tends to pull the valence electron away from the nucleus and therefore reduces the hyperfine interaction while the short-range repulsive force increases the interaction.
As shown in Fig.~\ref{figure-0}(b), a colliding molecule with relatively large electric polarizability, such as CH$_4$, usually has a dominant van der Waals attractive force, which therefore causes negative shifts of hyperfine interaction \cite{bernheim1969effects}. 
Whereas colliding molecules with relatively small electric polarizability as shown in Fig.~\ref{figure-0}(c), such as N$_2$, cause positive shifts \cite{bernheim1969effects}. 

\begin{figure}[htb]  
	\makeatletter
	\def\@captype{figure}
	\makeatother
	\includegraphics[scale=0.48]{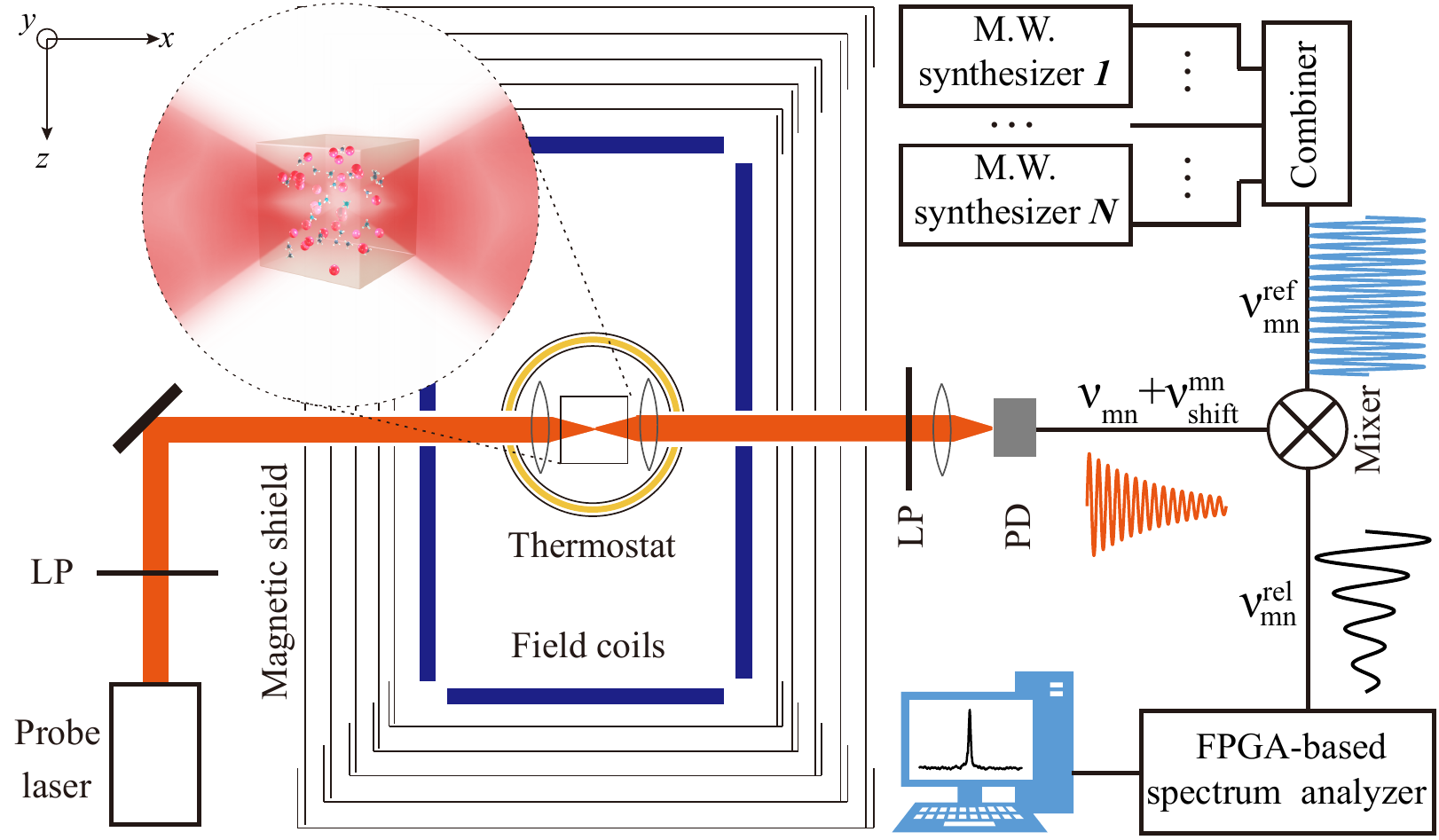}
	\caption{(color online). Schematic of measuring collision-induced spin noise. A vapor cell containing natural abundance Rb atoms and inert molecules is placed within an oven, and shielded inside a five-layer mu-metal shield. A 14~mW linearly probe laser is detuned 100~GHz from the Rb D2 optical transition at 780~nm. The laser beam is focused through the vapor cell with about 230~$\mu$m diameter. The intrinsic spin fluctuations of Rb spins $\delta S_x(t)$ impart small optical rotation fluctuations $\delta\theta\propto\delta S_x(t)$ on the probe laser~\cite{crooker2004spectroscopy}, detected by a linear polarizer (LP) and a photodetector (PD) with a bandwidth of 12.5~GHz. 
	The high-frequency raw noise signal is then mixed with an array of microwave (M.~W.) synthesizers to generate low-frequency signal (see text). 
	At last, the frequency down-converted noise signal is sent to a home-built FPGA-based spectrum analyzer~\cite{tong2020high}.
    }
	\label{figure-1}
\end{figure}

We now consider how to use spin-noise technique to measure collisions in a vapor cell.
The vapor cell comprises of natural abundance rubidium atoms and other colliding molecules to be studied.
As depicted in Fig.~\ref{figure-1},
linearly polarized light is focused through the vapor cell, the longitudinal spin polarizations of the atoms cause rotation of the light polarization. 
As the spin polarizations fluctuate randomly, the polarization rotation is in turn fluctuating.
The transient rotation can be measured with a high-bandwidth photodiode and analysed by a home-built field-programmable gate array (FPGA) based spectrum analyzer reported in our recent work \cite{tong2020high}.
According to the fluctuation-dissipation theorem \cite{kubo1966fluctuation,zapasskii2013spin,sinitsyn2016theory}, the spectrum of collision-induced spin noise can reveal the energy levels of colliding alkali atoms and thus collisional energy shifts.
This establishes a key bridge between collisional energy shifts and spin-noise spectrum
\begin{equation}
\begin{aligned}\label{downconvert}
S_{\alpha}({\nu})=&\sum\limits_{m,n}(\rho_m+\rho_n)\vert\bra{m}\textbf{S}\cdot\hat{\alpha}\ket{n}\vert^2
\frac{\gamma_{mn}}{(\nu-\nu_{mn}-\nu^{mn}_{\rm{shift}})^2+{\gamma_{mn}}^2},\\
\end{aligned}
\end{equation}
where $\rho_m$ is the occupation factor of level $m$ with eigenfrequency $\nu_m$.
$\hat{\alpha}$ is the unit vector along the probe laser propagation direction. 
$\nu_{mn}+\nu^{mn}_{\rm{shift}}=\nu_m-\nu_n+\nu^{mn}_{\rm{shift}}$ is the transition frequency of the  alkali atom, and is usually on the order of 1~GHz and 1~MHz for hyperfine splittings and Zeeman splittings, respectively.
The spin dephasing rate $\gamma_{mn}$, which is usually on the order of 1 kHz, determines the spectral resolution.

Comparing with traditional approaches to measuring collisional frequency shifts, such as coherent population trapping spectroscopy~\cite{robinson1960frequency,bernheim1969effects,vanier1974relaxation} and traditional microwave spectroscopy \cite{corsini2013hyperfine}, the spin-noise signals scale inversely with the probe volume \cite{crooker2004spectroscopy,zapasskii2013optical,sinitsyn2016theory}. 
Therefore, our technique is well suited to studying collisions at microscopic scales, which are now the focus of many studies in nanoscale sciences \cite{budker2019extreme}, such as local atom-surface collisions \cite{laliotis2014casimir,stern2014fano}.
Moreover, spin-noise techniques have been recently demonstrated in cold atoms with the capability of avoiding unwanted perturbations \cite{swar2021detection},
such as extra heating effects and atom loss caused by laser pumping \cite{kumar2018sorting}.
Thus, our technique provides a possible way for studying non-perturbative cold atom-molecule collisions \cite{shen2021refining}.

However, current spin-noise techniques are not suitable to measure collisions.
Existing spin-noise works mostly are limited to measure spin noise originated from Zeeman sublevels~\cite{crooker2004spectroscopy,li2013nonequilibrium,tang2020spin,lucivero2017correlation},
where the collisional frequency shifts are only on the order of 10$^{-2}$~Hz \cite{herman1968rare} and challenging to observe.
Moreover, although the hyperfine frequency shifts are on the order of 100~kHz, existing techniques suffer from a trade-off between bandwidth and spectral resolution.
Specifically,
the bandwidth is usually below 1~GHz because of the limited bandwidths of available balanced detectors and data acquisition cards (DAC) \cite{zapasskii2013spin,berski2013ultrahigh,cronenberger2016quantum,sinitsyn2016theory}.
Although some photodetectors and DAC can have gigahertz bandwidth, the applications of them to measuring gigahertz spin noise are still challenging since they require unrealistically high-speed real-time storage and large data processing \cite{crooker2010spin,tong2020high}.
Many ongoing efforts have been recently reported,
for example, with the use of ultrafast pair laser pulses~\cite{berski2013ultrahigh,zapasskii2013spin} or optical heterodyne detection ~\cite{cronenberger2016quantum},
but the state-of-the-art resolution is still limited to a few hundred~kilohertz \cite{cronenberger2016quantum}.


To address this difficulty, we introduce a frequency down-conversion technique as shown in Fig.~\ref{figure-1}, where the collision-induced spin-noise signal is converted from microwave to radio-frequency range.
Specifically, the raw spin-noise signals oscillating at $\nu_{mn}+\nu_{\rm{shift}}^{mn}$ are multiplied with multiple reference signals at $\nu^\textrm{ref}_{mn}$, which are close to $\nu_{mn}$, by a low-pass mixer to reserve the low-frequency signal at $\nu^\textrm{rel}_{mn}=\vert\nu_{mn}-\nu^{\textrm{ref}}_{mn}+\nu_{\rm{shift}}^{mn}\vert$ \cite{supplementary}.
Although $\nu_{mn}$ is on the order of 1~GHz, $\nu^\textrm{rel}_{mn}$ can be set as about 10~MHz.
Therefore the real-time detection of the down-converted spin-noise signal can be achieved with DAC with tens of megahertz bandwidth, significantly reducing the data amount by $\sim99\%$.
The phase noise of the reference signal causes the spectral broadening $\gamma_{\textrm{ref}}\sim 0.1$~Hz \cite{supplementary}, which is far smaller than $\gamma_{mn}$ and thus can be neglected. 

\begin{figure}[t]  
	\makeatletter
	\def\@captype{figure}
	\makeatother
\centering
	\includegraphics[scale=0.8]{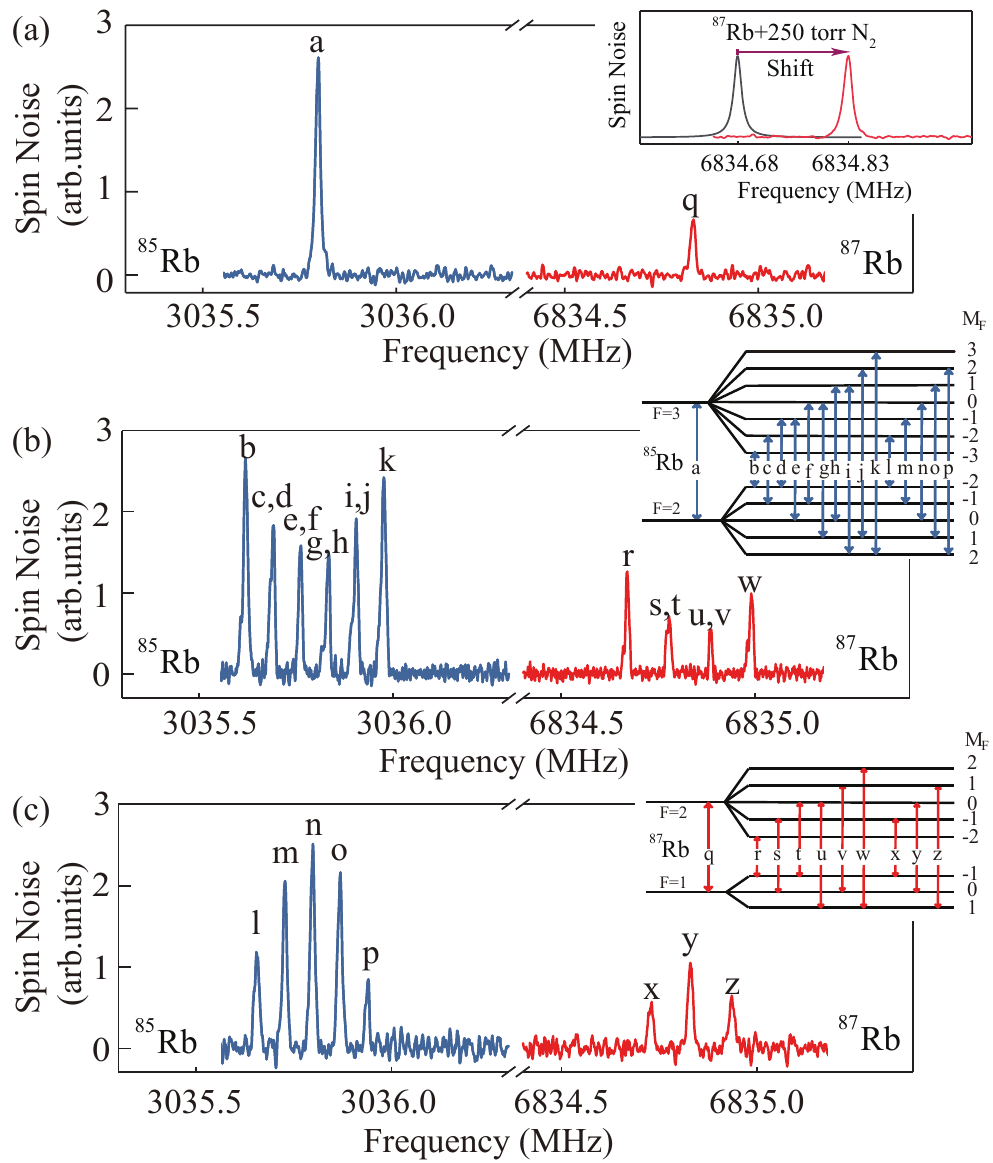}
	\caption{(color online). Collision-induced spin-noise experimental spectra.
    (a) Zero-field spin-noise spectra for $^{85}$Rb and $^{87}$Rb hyperfine levels. The inset shows a demonstration of $^{87}$Rb collision-induced frequency shift.
    (b) and (c), The measured spin-noise spectra in a 76-mG transverse and longitudinal magnetic fields, respectively. The corresponding spin-noise peaks correspond to those transitions between Zeeman sublevels belonging to different hyperfine sublevels (i.e., $\Delta$F=1, $\Delta$M$_F$=0, $\pm$1). The insets show the corresponding energy levels and allowed inter-hyperfine transitions of $^{85}$Rb and $^{87}$Rb.
    }
	\label{figure-2}
\end{figure}

Figure~\ref{figure-2} shows an example of the experimental noise spectra from Rb atoms colliding with N$_2$ molecules.
To simultaneously detect hyperfine transitions of two Rb isotopes, we use the array of two microwave synthesizers to down-convert the spin-noise signal. 
Figure \ref{figure-2}(a) shows such an experimental spin-noise spectrum of natural abundance Rb in zero applied magnetic field.
The blue and red noise spectra are from $^{85}$Rb and $^{87}$Rb atoms, yielding the experimental hyperfine splittings in collisions, i.e., 3035.79~MHz and 6834.83~MHz, respectively.
The theoretical hyperfine splittings are 3035.73 MHz and 6834.68 MHz for free $^{85}$Rb and $^{87}$Rb atoms, respectively.
A small frequency shift exists between the experimental and theoretical values, which is different for the two isotopes.
The observation of the small shift demonstrates the ability of our technique for measuring collision phenomena.
The observed effective spin dephasing rate is about $\gamma_{mn} \approx 7~\textrm{kHz}$, which is mainly due to the short transit time of atoms across the $\sim$230~$\mu$m focused probe laser beam \cite{crooker2004spectroscopy}.
In spite of this transit-time broadening, the resolution of our method is still at least two orders of magnitude better than existing gigahertz spin-noise techniques \cite{berski2013ultrahigh,cronenberger2016quantum}.

\begin{figure}[t]  
	\makeatletter
	\def\@captype{figure}
	\makeatother
\centering
	\includegraphics[scale=0.68]{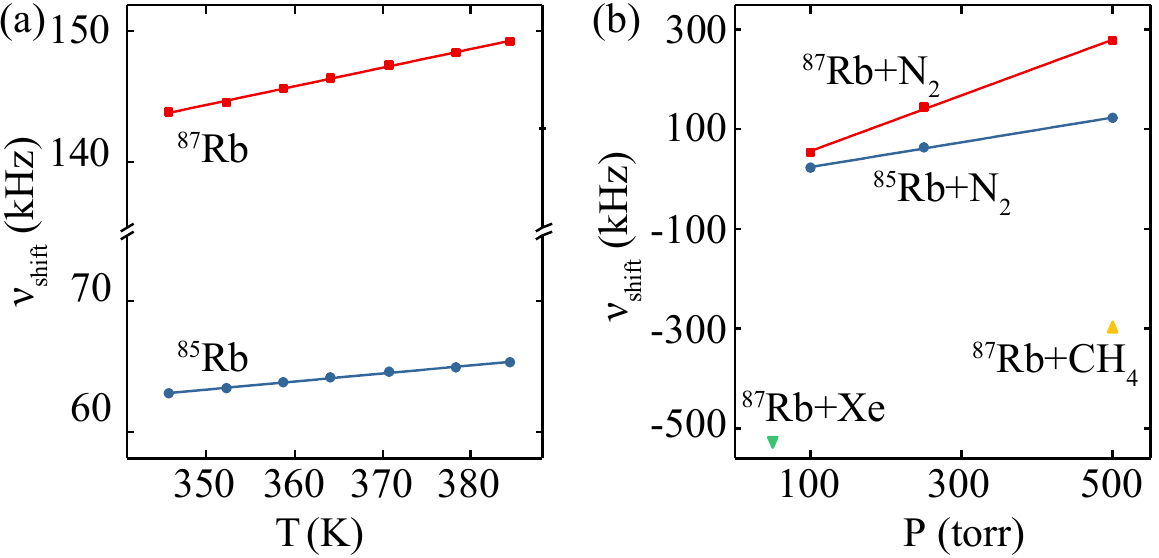}
	\caption{(color online).
	(a) Collisional shifts of Rb atoms versus vapor temperature in the presence of 250~torr N$_2$.
	The collisional shifts linearly increase with the temperature.
	(b) Collisional shifts of Rb atoms versus vapor pressure at $T_0$=337~K.
	The collisional shifts linearly increase with the pressure. In (a) and (b), error bars are within data points.}
	\label{figure-4}
\end{figure}

The collisional energy shifts of inter-hyperfine sublevels are measured.
To do this, a small magnetic field is applied.
Figure~\ref{figure-2}(b) and (c) show the spin-noise spectra in a transverse and longitudinal magnetic field ($\sim 76$~mG), respectively.
In these cases, the zero-field noise peak splits into resolvable multiplet, corresponding to the transitions with $\Delta F=1$, $\Delta M_F=\pm 1$ and $\Delta M_F=0$ in turn \cite{mihaila2006quantitative}, which is consistent with
Eq.~\eqref{downconvert}.
Every spin-noise peak equally shifts about 60~kHz and 150~kHz for $^{85}$Rb and $^{87}$Rb, respectively, which are equal to those in zero magnetic field.
This is because that the sublevels within the same hyperfine manifold shift equally, but those in different manifolds shift unequally.
Besides, the frequency difference between adjacent peaks is $\Delta\nu_{mn}\approx$ 71 (106)~kHz for $^{85}$Rb ($^{87}$Rb),
which is approximately equal to $\Delta\nu_{mn}\approx2 g_F\mu_BB$ with g-factor $\vert g_F\vert\cong g_S/(2I+1)$.
We observe no significant collisional shifts of Zeeman sublevels.
This verifies that the collisional shift of Zeeman sublevels is significantly smaller than that of inter-hyperfine sublevels.


We would like to emphasize the difference between our technique and traditional spectrum analyzers.
Due to no need of high-bandwidth DAC, the spectrum analyzers employing swept local oscillators are usually adopted to measure the signals at gigahertz range, however they ignore most available data  \cite{romer2007spin}.
For example, measuring a spectrum with 10~MHz window and 1~kHz resolution effectively uses only $\sim$0.01$\%$ of the data \cite{crooker2010spin}.
Contrastly, our approach avoids this by fast-Fourier transforming of the down-converted data in a FPGA \cite{tong2020high},
realizing the 100$\%$ data utilization and therefore reducing the measurement time by more than three orders of magnitude.
Moreover, with only one frequency component detected each time, the spectrum analyzers can not extract correlations between different frequencies \cite{roy2015cross},
which are useful to study many-body interactions
in correlated spin systems \cite{sinitsyn2016theory,li2016universality}.
To overcome this, we employ multiple reference signals to simultaneously measure the spin noise at the corresponding frequencies, as demonstrated in Fig.~\ref{figure-2}.
In addition, the spectrum analyzers are not suitable for studying higher-order spin-noise correlations that have multi-time moments \cite{li2016universality,wang2019characterization}. 
With the capability of real-time measurement, our technique has the natural potential of measuring higher-order cumulants.

\emph{Measurement of collision parameters}.-With the use of collision-induced spin noise spectra,
the collision parameters can be precisely determined.
As shown in Fig.~\ref{figure-4}(a), our experiments are performed in a relatively small temperature range, and the measured collisional shift is nearly linear to the vapor temperature.
In this case, the collisional shift can be expanded to the first order of $(T-T_0)$.
According to the Eq.~\ref{shift1}, the collisional shift is proportional to the number density of colliding molecules, which is confirmed by our experiments shown in Fig.~\ref{figure-4}(b).
Here we change the number density via changing the pressure of the colliding molecules at a constant temperature $T_0$.
As a result, the collisional shift in Eq.~\eqref{shift1} can be approximated as
\begin{equation}
\begin{aligned}\label{shift4}
\nu_{\text{shift}} \approx P_0[\beta+\delta(T-T_0)]+O\left((T-T_0)^2\right),\\
\end{aligned}
\end{equation}
where $P_0$ and $T_0$ are the reference pressure and temperature of the vapor cell, respectively, and are calibrated through absorption spectra of Rb atoms \cite{supplementary}.
$\beta$ is the pressure-dependent coefficient and measured to be 559~Hz/torr and 249~Hz/torr for $^{87}$Rb$-$N$_2$ and $^{85}$Rb$-$N$_2$ pairs, respectively.
$\delta$ is the temperature-dependent coefficient and measured to be 0.57~Hz/(K$\cdot$torr) and 0.25~Hz/(K$\cdot$torr) for $^{87}$Rb$-$N$_2$ and $^{85}$Rb$-$N$_2$ pairs, respectively.
We find the ratio between the pressure-dependent or temperature-dependent shifts of two Rb isotopes are 2.28 and 2.24, respectively, which are close to their hyperfine splitting ratio (2.25) \cite{vanier1974relaxation,budker2005microwave}.

We now consider the intermolecular potential as the Lennard-Jones potential in Eq.~\eqref{shift2}, which has been widely studied in analyzing gas collisions \cite{kaplan2006intermolecular,robinson1960frequency}.
Based on Eq.~\eqref{shift1}, we expand the shift using Eq.~\eqref{shift4},
and get the relation between pressure-dependent coefficient $\beta$ and temperature-dependent coefficient $\delta$ with parameters $\{\epsilon_1,  \epsilon_2, \sigma_1, \sigma_2\}$ \cite{supplementary}. 
Combined with other two theoretical formulas, simultaneous analysis of $\beta(\epsilon_1,  \epsilon_2, \sigma_1, \sigma_2)$ and $\delta(\epsilon_1,  \epsilon_2, \sigma_1, \sigma_2)$ can finally derive the four collision parameters \cite{supplementary}.
Specifically, for the colliding $^{87}$Rb$-$N$_2$ pair, we obtain the collision diameter between Rb atoms and colliding N$_2$ molecules $\sigma_1\approx4.19$~$\mathring{\textrm{A}}$, the well depth $\epsilon_1\approx7.6$~meV, which are in good agreement with the theoretical results \cite{robinson1960frequency}.
Moreover, we find that the collision parameters $\{\epsilon_1,  \epsilon_2, \sigma_1, \sigma_2\}$ for $^{85}$Rb-N$_2$ pairs are the same with those of $^{87}$Rb$-$N$_2$ pairs,
yielding that collision parameters are usually independent of the nuclear structure of Rb isotopes.

We also test the feasibility of our technique to investigate $^{87}$Rb atoms colliding with different atoms and molecules, for example, CH$_4$ and Xe.
The frequency shifts induced by N$_2$ molecules with relatively small electric polarizability are positive [see Fig.~\ref{figure-0}(c)], yielding that the Pauli exclusion is dominant for Rb-N$_2$ pairs \cite{bernheim1969effects}. 
In contrast, when Rb atoms collide with molecules that have relatively large electric polarizability [see Fig.~\ref{figure-0}(b)], such as Xe and CH$_4$ [see Fig.~\ref{figure-4}(b)], the van der Waals interaction dominates \cite{bernheim1969effects}.
Our result clearly shows the dominant potential type for different colliding pairs via the sign of frequency shifts.
This suggests a convenient way to analyse the information of colliding molecules, such as molecular electric polarizability.


\emph{Conclusion}.-In this work, we have proposed and demonstrated a new spin-noise technique.
This technique is capable of characterizing collision phenomena in alkali atoms and other molecules, providing important information of the intermolecular potential,
collision diameter, potential-well depth and dominant interaction type.
In contrast to frequently used scattering approaches that require ultra-high vacuum systems, our technique employs much simpler and lower-cost apparatus but still with high precision.
The present approach can be extended to investigate complicated intermolecular potentials, for example modified Lennard-Jones potential~\cite{robinson1960frequency,karman20182} and Buckingham potential \cite{kaplan2006intermolecular}.
To measure such potentials, we can measure the collision-induced spin noise under a large temperature range,
then obtain high-order coefficients of $T^n$ term of $\nu_{\textrm{shift}}$, and in turn extract the collision parameters.
Although our work focuses on the binary collisions, it can be extended to many-body collisions, such as the three-body collisions in the van der Waals molecules \cite{gong2008nonlinear}.
Moreover,
our technique significantly improves the experimental efficiencies in ongoing experimental efforts to measure quantum noise with part-per-million resolution over tens of gigahertz frequency range,
opening a feasible route towards quantum noise-based applications, for example, non-perturbative structural analysis for diverse spin systems \cite{crooker2004spectroscopy,cronenberger2015atomic}, determining the fundamental precision of microwave quantum devices \cite{pedrozo2020entanglement} and the degree of squeezing and entanglement \cite{kong2020measurement,bao2020spin}, and researching many-body phase transitions \cite{eckert2008quantum,bruun2009probing,chen2014faraday,wang2019characterization}.



~\

We thank Dmitry Budker and Kaifeng Zhao for valuable discussions.
This work was supported by National Key Research and Development Program of China (grant no.~2018YFA0306600), National Natural Science Foundation of China (grants nos. 11661161018, 11927811, 12004371), Anhui Initiative in Quantum Information Technologies (grant no.~AHY050000), the Hong Kong RGC/NSFC Joint Research Scheme Project N\_CUHK403/16, and USTC Research Funds of the Double First-Class Initiative (grant no. YD3540002002).


\bibliographystyle{naturemag}
\bibliography{noise}

\end{document}